\documentclass[print]{revtex4}
\usepackage{amsmath,amssymb}
\usepackage{graphicx}

\draft

\begin{document}

\title{A Scheme of Generating and Spatially Separating Two-Component Entangled
Atom Lasers}
\author{Xiong-Jun Liu$^{a,c}$\footnote{Electronic address:
phylx@nus.edu.sg}, Hui Jing$^{b}$, Xin Liu$^{c}$, Ming-Sheng
Zhan$^{b}$ and Mo-Lin Ge$^{c}$} \affiliation{ a. Department of
Physics, National University of Singapore, 10 Kent Ridge Crescent,
Singapore 119260, Singapore \\
b. State Key Laboratory of Magnetic Resonance and
Atomic and Molecular Physics,\\
 Wuhan Institute of Physics and Mathematics, CAS, Wuhan 430071, P. R.
 China\\c. Theoretical Physics Division, Nankai Institute of
Mathematics,Nankai University, Tianjin 300071, P.R.China}

\begin{abstract}
Entanglement of remote atom lasers is obtained via quantum state
transfer technique from lights to matter waves in a five-level
$M$-type system. The considered atom-atom collisions can yield an
effective Kerr susceptibility for this system and lead to the
self- and cross- phase modulation between the two output atom
lasers. This effect results in generation of entangled states of
output fields. Particularly, under different conditions of
space-dependent control fields, the entanglement of atom lasers
and of atom-light fields can be obtained, respectively.
Furthermore, based on the Bell-state measurement, an useful scheme
is proposed to spatially separate the generated entangled atom
lasers.

\end{abstract}
\maketitle

\indent Quantum entanglement has always attracted great interest
as it is one of the key differences between quantum and classical
physics. Since it can be exploited for various novel applications
such as quantum computation and precision measurements, there has
been a continuing effort to engineer the robust quantum entangled
states in different systems. Since the experimental realization of
Bose-Einstein Condensation (BEC) in dilute atomic clouds in 1995,
much efforts have been taken in preparing a continuous atom laser
\cite{atom laser} and exploring its potential applications in,
e.g., gravity measurements through atom interferometry
\cite{interferometer}. The method for creating two correlated
matter waves has been proposed via four-wave mixing using BECs
several years ago \cite{deng}, and the large amplification of the
generated correlated matter waves was also achieved
\cite{ketterle}. Entanglement between the generated matter waves
is possible when consider the coherent collisions between
condensate atoms, which has been observed before \cite{coherence}.

Here we propose a scheme to generate and spatially separate
entangled atom lasers from a five-level $M$-type system, with
coherent collisions between atoms considered. This technique is
based on the physical mechanism of Electromagnetically Induced
Transparency (EIT) \cite{1} which has attracted much attention in
both experimental and theoretical aspects \cite{2,3,4,wu},
especially for the rapid developments of quantum memory technique
\cite{5}, i.e., transferring the quantum states of photon
wave-packet to collective Raman excitations in a loss-free and
reversible manner. The quantum state transfer technique via EIT
provides a new optical technique to generate continuous atom laser
with extra quantum states \cite{6,liu}. Very recently, considering
the nonlinear effect in the EIT quantum state transfer process,
many intriguing applications were discovered by a series
publications, such as the solitons formed by dark-state polaritons
in the EIT Kerr medium \cite{4}, generation of quantum phase gate
for photons \cite{five} and nonclassical soliton atom laser
\cite{sal} by considering the coherent atom-atom collisions.

In the following, firstly we investigate how to generate
two-component entangled atom lasers by considering the atom-atom
collisions in the quantum state transfer technique from lights to
matter waves in a five-level $M$-type system. The considered
atom-atom collisions can yield an effective Kerr susceptibility
for the probe lights and lead to the self- and cross-phase
modulation between the two output atom lasers. This effect is
useful for generation of entangled states. Then, we propose a
scheme to spatially separate the generated entangled atom lasers
via entanglement swapping technique \cite{Bell}.

The system we considered is shown in Fig.1. A beam of five-level
$M$ type atoms moving in the $z$ direction interact with two
quantized probe and two classical control Stokes fields
\cite{6,liu}, and the former fields are taken to be much weaker
than the later ones. Atoms in different internal states are
described by five bosonic fields $\hat\Psi_{\mu}({\bf r},t)
(\mu=b,q_1,q_2,e_1,e_2)$. The two Stokes fields coupling the
transitions from the state $|q_j\rangle$ to excited one
$|e_j\rangle$ $(j=1,2)$ can be described by the Rabi-frequencies
$\Omega_{c_j}=\Omega_{0j}(z)e^{-i\omega_{s}(t-z/c_j)}$,
respectively, with $\Omega_{0j}$ being taken as real, and $c_j$
denoting the phase velocities projected onto the $z$ axis. The
quantized probe fields coupling the transitions from the state
$|b\rangle$ to $|e_j\rangle$ are characterized by the
dimensionless positive frequency components $\hat
E_j(z,t)=\hat{\cal E}_j(z,t)e^{-i\omega_{p_j}(t-z/c)}$. The
Heisenberg equations for bosonic field operators under the
$s$-wave approximation are governed by \cite{five}
\begin{eqnarray}\label{eqn:1}
i\hbar\frac{\partial\hat\Psi_b}{\partial
t}&=&[-\frac{\hbar^2}{2m}\nabla^2+\hbar\omega_b+V(\bold
r)+\mu_{b}\hat\Psi_b^{\dag}\hat\Psi_b+\sum_{k=1,2}\mu_{bk}\hat\Psi_{q_k}^{\dag}\hat\Psi_{q_k}]\hat\Psi_b\nonumber\\
&& \ \ \ \ \ \ \ \ \ \ \ \ +\hbar g\hat E_1^{\dag}\Psi_{e_1}+\hbar
g\hat E_2^{\dag}\hat\Psi_{e_2},
\end{eqnarray}
\begin{eqnarray}\label{eqn:2}
i\hbar\frac{\partial\hat\Psi_{q_j}}{\partial
t}&=&[-\frac{\hbar^2}{2m}\nabla^2+\hbar\omega_{q_j}+V(\bold
r)]\hat\Psi_{q_j}+[\mu_{bj}\hat\Psi_b^{\dag}\hat\Psi_b+\sum_{k=1,2}\mu_{jk}\hat\Psi_{q_k}^{\dag}\hat\Psi_{q_k}
]\hat\Psi_{q_j}\nonumber\\
&& \ \ \ \ \ \ \ \ \ \ \ \ +\hbar \Omega_{c_j}^*\hat\Psi_{e_j},
\end{eqnarray}
\begin{eqnarray}\label{eqn:4}
i\hbar\frac{\partial\hat\Psi_{e_j}}{\partial
t}=[-\frac{\hbar^2}{2m}\nabla^2+\hbar\omega_{e_j}+V(\bold
r)]\hat\Psi_{e_j}+\hbar\Omega_{c_j}\hat\Psi_{q_j}+\hbar g\hat
E_j\hat\Psi_b,
\end{eqnarray}
where $j=1,2$, $V(\bold r) $ is the external trap potential, the
scattering length $a_{ij}$ characterizes the atom-atom
interactions via $\mu_{ij}=4\pi\hbar^2a_{ij}/m$ $i,j=b,1,2$ and
$\omega_j$'s are the frequencies corresponding to the electronic
energy levels. Since in present discussion almost no atoms occupy
the excited states $|e_1\rangle$ or $|e_2\rangle$ in the
dark-state condition that is fulfilled in EIT technique, the decay
from the excited states and the collisions between the excited
states and lower states can be safely neglected. The motion
equations for the two quantized probe fields read
\begin{eqnarray}\label{eqn:elm1}
(\frac{\partial}{\partial t}+c\frac{\partial}{\partial z})\hat
E_{1,2}(z,t)=ig\int
d^2r_{\perp}\hat\Psi_b^{\dag}\hat\Psi_{e_{1,2}}
\end{eqnarray}
where $r_{\perp}$ represents the transverse coordinate
(perpendicular to the $z$ axis).

The transverse motion (perpendicular to the $z$ axis) of the beam
is confined by the transverse trapping potential. Within the
adiabatic condition the transverse motion can be restricted to the
lowest transverse eigen-state \cite{adiabatic}. For this the
bosonic fields can be cast into two parts \cite{form}, i.e. the
transverse part and the longitudinal part (along $z$ axis). On the
other hand, the EIT quantum state transfer process requires that
the Rabi-frequencies of the Stocks control fields vary
sufficiently slowly on the $z$ coordinate, and then all the
bosonic fields' amplitudes vary slowly with time and $z$
coordinate during the propagation, i.e., we can further
conveniently introduce the slowly-varying amplitudes and a
decomposition into velocity class \cite{6,liu} for the
longitudinal part of each bosonic field. Therefore we rewrite the
atomic fields as: $
\hat\Psi_b=\sum_l\psi(\vec{r}_{\perp};z)\hat\Phi^l_b(z,t)e^{i(k_lz-\omega_lt)},
\hat\Psi_{e_j}=\sum_l\psi(\vec{r}_{\perp};z)\hat\Phi^l_{e_j}(z,t)e^{i[(k_l+k_{p_j})z-(\omega_l+\omega_{p_j})t}$
and $
\hat\Psi_{q_j}=\sum_l\psi(\vec{r}_{\perp};z)\hat\Phi^l_{q_j}(z,t)e^{i[(k_l+k_{p_j}-k_{s_j})z
-(\omega_l+\omega_{p_j}-\omega_{c_j})t}$, where $j=1,2$,
$\hbar\omega_l=\hbar^2k^2_l/2m$ is the corresponding kinetic
energy in the $l$th velocity class, $k_{p_j}$ and $k_{s_j}$
$(j=1,2)$ are respectively the vector projections of the probe and
Stokes fields to the $z$ axis. The atoms have a narrow velocity
distribution around $v_0=\hbar{k_0}/m$ with $k_0\gg|k_p-k_s|$, and
all fields are assumed to be in resonance for the central velocity
class. $\hat\Phi^l_{\mu}(z,t)$ are the slowly-varying amplitudes
that describe the motion of bosonic fields along $z$ axis, and
$\psi(\vec{r}_{\perp};z)$ describes the equilibrium wave function
(normalized to unity) in the transverse direction, i.e. $\int
d^2r_{\perp}\psi^*(\vec{r}_{\perp};z)\psi(\vec{r}_{\perp};z)=1$.
In our case the transverse trapping potential part
$V(\vec{r}_\perp)$ is set to be independent of $z$ and the
effective longitudinal potential can then be taken to be zero (see
ref. \cite{form,Leboeuf}) (This requirement is similar to that in
previous publications \cite{6,liu}). Thus the equations of motion
for the field operators can be recast into
\begin{eqnarray}\label{eqn:7}
\bigr(\frac{\partial}{\partial t}+\frac{\hbar
k_l}{2m}\frac{\partial}{\partial
z}\bigr)\Phi_b^l=-i\sum_{j=1,2}g\hat{\cal
E}_j^{\dag}\Phi^l_{e_j}-i\sum_l\bigr(\mu_b\Phi_b^{\dag
l}\Phi_b^{l}+\sum_{j=1,2}\mu_{bj}\Phi_{q_j}^{\dag
l}\Phi_{q_j}^{l}\bigr)\Phi_b^l,
\end{eqnarray}
\begin{eqnarray}\label{eqn:8}
\bigr(\frac{\partial}{\partial t}+\frac{\hbar
k_l}{2m}\frac{\partial}{\partial
z}\bigr)\Phi_{q_j}^l=-i\delta_j^l\Phi_{q_j}^l-i\Omega_{0j}\Phi^l_{e_j}-i\sum_{k=1,2}\sum_l\bigr(\mu_{bk}\Phi_b^{\dag
l}\Phi_b^{l}+\mu_{jk}\Phi_{q_k}^{\dag
l}\Phi_{q_k}^{l}\bigr)\Phi_{q_j}^l,
\end{eqnarray}
\begin{eqnarray}\label{eqn:9}
\bigr(\frac{\partial}{\partial t}+\frac{\hbar
(k_l+k_{p_j})}{2m}\frac{\partial}{\partial
z}\bigr)\Phi_{e_j}^l=-\Delta_j^l\Phi_{e_j}^l-ig\hat{\cal
E}_j\Phi^l_{b}-i\Omega_{0j}\Phi^l_{q_j}.
\end{eqnarray}
and the two probe lights:
\begin{eqnarray}\label{eqn:prob1}
(\frac{\partial}{\partial t}+c\frac{\partial}{\partial
z})\hat{\cal E}_j(z,t)=-ig\sum_l\hat\Phi_b^{\dag
l}\hat\Phi^l_{e_j},
\end{eqnarray}
where
$\Delta^l_j\approx\hbar{k_l}k_{p_j}/m+(\omega_{e_jb}-\omega_{p_j})$
and
$\delta^l_j\approx\hbar{k_l}(k_{p_j}-k_{s_j})/m+(\omega_{q_jb}-\omega_{p_j}-\omega_{c_j})$
are the single and two-photon detunings, respectively.

In order to solve the above equations of the matter-field
operators, we shall consider the weak-probe-field approximation.
In this case, consider a stationary input of atoms in state
$|b\rangle$ and in zero order of the probe fields, the depletion
of the atoms in the state $|b\rangle$ is neglected. Therefore we
can make the replacement $\sum_l\Phi_b^{\dag
l}\Phi_b^{l}\rightarrow n$ \cite{6,liu}, where $n$ the constant
total atomic density. Next, we shall consider the perfect one- and
two-photon resonance case, i.e. $\Delta_j^l=0$ and
$\delta_j^l+\mu_{bj}n=0$ so that we can apply an adiabatic
approximation (the validity of this approximation will be
discussed later) for the longitudinal motion of the fields and
find: $
\Phi^l_{e_j}=-\frac{1}{\Omega_{0j}}\sum_{k=1,2}\sum_l\mu_{jk}\Phi^{\dag
l}_{q_k}\Phi^l_{q_k}\Phi^l_{q_j}+\frac{i}{\Omega_{0j}}\bigr(\frac{\partial}{\partial
t}+\frac{\hbar k_l}{2m}\frac{\partial}{\partial
z}\bigr)\Phi^l_{q_j}$ and $$
\Phi^l_{q_1}\approx-\frac{g}{\Omega_{01}}\hat{\cal E}_1\Phi^l_b, \
\ \ \Phi^l_{q_2}\approx-\frac{g}{\Omega_{02}}\hat{\cal
E}_2\Phi^l_b$$ with $\Phi^l_b=\sqrt{n}\xi_le^{-i(k_lz-\omega_lt)}$
and $\sum_l\xi_l=1$ \cite{6,liu}. Substitute these results into
the above propagation equations of the two probe yields
\begin{eqnarray}\label{eqn:prob2}
&&\bigr[(1+\frac{g^2n}{\Omega^2_{0j}(z)})\frac{\partial}{\partial{t}}
+c(1+\frac{g^2n}{\Omega^2_{0j}(z)}\frac{v_0}{c})
\frac{\partial}{\partial{z}}\bigl]\hat{\cal E}_j(z,t)\nonumber\\
&&+i\sum_{k=1,2}\frac{g^4n^2}{\Omega^4_{ok}(z)}\mu_{jk}\hat{\cal
E}_k^{\dag}\hat{\cal E}_k\hat{\cal E}_j(z,t)
=\frac{g^2n}{\Omega^2_{0j}(z)}v_0(\frac{\partial}{\partial{z}}
\ln\Omega_{0j}(z))\hat{\cal E}_j(z,t).
\end{eqnarray}
where $j=1,2$. The last part of the l.h.s of the this equation
indicates an effective Kerr nonlinear interaction and the r.h.s.
describes a reduction (enhancement) due to stimulated Raman
adiabatic passage in two spatially varying Stokes fields for
$v_0\neq0$. With the definitions of the mixing angles $\theta_j$
by $\tan^2\theta(z)\equiv\frac{g^2n}{\Omega_0^2}\frac{v_0}{c}$,
one finds the solutions for the probe fields:
\begin{eqnarray}\label{eqn:16}
\hat{\cal E}_j(z,t)&=&\exp\bigl[-i\sum_{k=1,2}\mu_{jk}\hat{\cal
E}_k^{\dag}(0,T_j)\hat{\cal E}_k(0,T_j)
\int_0^z|\frac{\cos\theta_j(\xi)}{\cos\theta_k(0)}|^2\frac{g^4n^2}
{\Omega^2_{0k}(\xi)(\Omega^2_{0j}(\xi)+g^2n\frac{v_0}{c})}d\xi\bigl]\times\nonumber\\
&&\times\frac{\cos\theta_j(z)}{\cos\theta_j(0)}\hat{\cal
E}_j\bigl(0,T_j\bigl),
\end{eqnarray}
where $T_j=t-\int dz'/V^{j}_{g_r}$ with the group velocity
$V^{j}_{g_r}=c(1+\frac{g^2n}{\Omega_{0j}^2}\frac{v_0}{c})/(1+\frac{g^2n}{\Omega_{0j}^2})$.
The above formula clearly shows that the atom-atom collisions
leads to self- and cross-phase modulation, and the additional
phase is dependent on the Rabi-frequency $\Omega_0(z)$ of the
stokes fields and the velocity $v_0$ of atomic beam. Assuming at
the entrance point $\theta=0$, i.e. the both control fields are
much stronger at the entrance point, one finds that the output two
corresponding atom lasers read
\begin{eqnarray}\label{eqn:out1}
\hat\Phi_{q_{1,2}}(z=L,t)=\sqrt{\frac{c}{v_0}}\exp\bigl[-i(\widehat{SPM}_{1,2}
+\widehat{CPM}_{1,2})\bigl]\sin\theta_{1,2}(L)\hat{\cal
E}_{1,2}(0,t),
\end{eqnarray}
where $\hat\Phi_{q_j}=\hat\Psi_{q_j}e^{-i[(k_{p_j}-k_{s_j})z
-(\omega_{p_j}-\omega_{c_j})t]}$ and the self-phase-modulation
$\widehat{SPM_j}$ and the cross-phase-modulation $\widehat{CPM_j}$
can easily be obtained from the eqs. (\ref{eqn:16})
\begin{eqnarray}\label{eqn:phase1}
\widehat{SPM_j}(z,t)=\mu_{jj}\hat{\cal E}_j^{\dag}(0,T)\hat{\cal
E}_j(0,T)\int_0^z\cos\theta_j^2(\xi)\sum_{k=1,2}\frac{(1-\Theta_{jk})g^4n^2}{
\Omega^2_{0k}(\xi)(\Omega^2_{0j}(\xi)+g^2n\frac{v_0}{c})}d\xi,
\end{eqnarray}
\begin{eqnarray}\label{eqn:phase2}
\widehat{CPM_j}(z,t)=\sum_{j,k=1,2}\mu_{jk}(1-\Theta_{jk})\hat{\cal
E}_k^{\dag}(0,T)\hat{\cal
E}_k(0,T)\int_0^z\cos\theta^2_j(\xi)\frac{g^4n^2}{\Omega^2_{0k}(\xi)(\Omega^2_{0j}(\xi)+g^2n\frac{v_0}{c})}d\xi,
\end{eqnarray}
where the symbol $\Theta_{jk}$ is defined by $\Theta_{jk}=1$ for
$j=k$, and $\Theta_{jk}=1$ for $j\neq k$. The result
(\ref{eqn:out1}) shows that when $\theta_j(L)=\pi/2$, i.e. the
Rabi-frequency of control field $\Omega_j$ decreases to be much
small at the output point, quantum states of probe light can be
fully transferred to the corresponding output atomic laser.

The self-phase-modulation of the output states may lead to
frequency-chirp effect and the cross-phase-modulation may lead to
entanglement between the output states, which can be studied using
proper space-varying control fields. Firstly, we consider the case
shown in Fig.2(a), i.e. the mixing angles satisfy:
\begin{eqnarray}\label{eqn:out2}
\theta_1(L)=\theta_2(L)=\pi/2\nonumber.
\end{eqnarray}
In this case, the initial quantum states of probe lights are fully
transferred to atom lasers associated with nontrivial phase
shifts. For example, for the most ``classical" states of two input
probe lights $|\alpha\rangle\otimes|\beta\rangle$, where
$|\alpha\rangle$ and $|\beta\rangle$ are single-mode coherent
states, when $\mu_{11}=\mu_{22}=2\mu_{12}(=2\mu_{21})$, and the
phase shift
\begin{eqnarray}\label{eqn:phase3}
\mu_{12}\int_0^L|\cos\theta_j(\xi)|^2\frac{g^4n^2}{\Omega^2_{0k}(\xi)
(\Omega^2_{0j}(\xi)+g^2n\frac{v_0}{c})}d\xi=\pi,
\end{eqnarray}
with $j=1$ for $k=2$ and $j=2$ for $k=1$, the output state of the
two atom lasers can be verified as
$|\Psi\rangle_{atom}=\frac{1}{2}(|\bar\alpha,\bar\beta\rangle+
|-\bar\alpha,\bar\beta\rangle+|\bar\alpha,-\bar\beta\rangle-
|-\bar\alpha,-\bar\beta\rangle)$ with
$\bar\alpha=\sqrt{c/v_0}\alpha, \bar\beta=\sqrt{c/v_0}\beta$. This
is an entangled superposition of macroscopically distinguishable
states. Furthermore, if we consider the case shown in Fig.2(b),
i.e. the mixing angle $\theta_1(L)=\pi/2$ whereas $\theta_2(L)=0$,
only the atom laser $\hat\Phi_{q_1}$ is generated and the second
probe light is emitted out. In this way, under the condition of
Eq. (\ref{eqn:phase3}), we find the input state
$|\alpha\rangle\otimes|\beta\rangle$ finally evolves into
\begin{eqnarray}\label{eqn:out3}
|\Psi\rangle_{a-l}=\frac{1}{2}|\bar\alpha\rangle_{A}\bigr(|\beta\rangle+|-\beta\rangle\bigr)_{L}
+\frac{1}{2}|-\bar\alpha\rangle_{A}\bigr(|\beta\rangle-|-\beta\rangle\bigr)_{L},
\end{eqnarray}
where $A$, $L$ represent the output $atom$ $laser$ and $probe$
$light$, respectively. Eq. (\ref{eqn:out3}) shows that the
entangled state between atom laser and photons can be readily
obtained with our model. Since light speed is much larger than
that of the generated atom laser, the above entanglement is
between two distant qubits. This result is useful for quantum
teleportation.

However, a challenge is still left for the two entangled atom
lasers: Since they propagate in the same direction, generally the
entanglement of the output atom lasers $|\Psi\rangle_{atom}$ is
local, whereas for practical applications we should separate them
spatially. This issue can be studied with entanglement swapping
technique \cite{Bell,Bell2}.

The schematic set-up is shown in Fig. 3, by which we demonstrate
how to spatially separate the two-component entangled atom lasers.
M is a semitransparent mirror splitter, through which the two
input probe pulses $\hat E_1(z,t)$ and $\hat E_2(z',t)$ are split
into four pulses with identical expectative intensities, i.e.
$\hat E'_1(z,t)$, $\hat E'_2(z,t)$, $\hat E''_1(z't)$ and $\hat
E''_2(z',t)$ with their amplitudes equal each other. After these
splitters, the four pulses enter the two channels (channel 1 and
channel 2), respectively. Also consider the input coherent states
of two probe lights $|\alpha\rangle\otimes|\beta\rangle$, together
with the conditions shown in Fig. 2(c) and eq. (\ref{eqn:phase3}),
we find the output states from the two channels
\begin{eqnarray}\label{eqn:entangled}
|\Psi\rangle&=&|\Psi\rangle_1\otimes|\Psi\rangle_2\nonumber\\
&=&\frac{1}{4}\bigl(|\frac{\bar\alpha}{\sqrt{2}}\rangle_{1A}\otimes(|\frac{\beta}{\sqrt{2}}\rangle+|\frac{-\beta}{\sqrt{2}}\rangle)_{1L}
+|\frac{-\bar\alpha}{\sqrt{2}}\rangle_{1A}\otimes(|\frac{\beta}{\sqrt{2}}\rangle-|\frac{-\beta}{\sqrt{2}}\rangle)_{1L}\bigl)\otimes\nonumber\\
&&\otimes\bigl(|\frac{\alpha}{\sqrt{2}}\rangle_{2L}\otimes(|\frac{\bar\beta}{\sqrt{2}}\rangle+|\frac{-\bar\beta}{\sqrt{2}}\rangle)_{2A}
+|\frac{-\alpha}{\sqrt{2}}\rangle_{2L}\otimes(|\frac{\bar\beta}{\sqrt{2}}\rangle-|\frac{-\bar\beta}{\sqrt{2}}\rangle)_{2A}\bigl)\\
&=&\frac{1}{4}\bigl(\sqrt{N_+N'_+}|\frac{\bar\alpha}{\sqrt{2}}\rangle_{1A}\otimes|\frac{\bar\beta}{\sqrt{2}}\rangle_{2A}\otimes|+\rangle_{1L}\otimes|+\rangle_{2L}
+\sqrt{N_+N'_-}|\frac{\bar\alpha}{\sqrt{2}}\rangle_{1A}\otimes|\frac{-\bar\beta}{\sqrt{2}}\rangle_{2A}\otimes|+\rangle_{1L}\otimes|-\rangle_{2L}\nonumber\\
&&+\sqrt{N'_+N_-}|\frac{-\bar\alpha}{\sqrt{2}}\rangle_{1A}\otimes|\frac{\bar\beta}{\sqrt{2}}\rangle_{2A}\otimes|-\rangle_{1L}\otimes|
+\rangle_{2L}+\sqrt{N_-N'_-}|\frac{-\bar\alpha}{\sqrt{2}}\rangle_{1A}\otimes|\frac{-\bar\beta}{\sqrt{2}}\rangle_{2A}\otimes|-\rangle_{1L}\otimes|-\rangle_{2L}\bigl)\nonumber,
\end{eqnarray}
where $|\Psi\rangle_1$ and $|\Psi\rangle_2$ represent the output
states from channel 1 and channel 2. The orthogonal bases are
defined as
$|\pm\rangle_{1L}=\frac{1}{\sqrt{N_{\pm}}}\bigl(|\frac{\beta}{\sqrt{2}}\rangle_{1L}\pm|\frac{-\beta}{\sqrt{2}}\rangle_{1L}\bigl)$
and
$|\pm\rangle_{2L}=\frac{1}{\sqrt{N'_{\pm}}}\bigl(|\frac{\alpha}{\sqrt{2}}\rangle_{2L}\pm|\frac{-\alpha}{\sqrt{2}}\rangle_{2L}\bigl)$
which correspond to quantum states of light fields output from
channel $1$ and channel $2$, respectively, with the normalized
factors $N_{\pm}=2\pm2e^{-|\alpha|^2}$ and
$N'_{\pm}=2\pm2e^{-|\beta|^2}$. The factorizable state in
eq.(\ref{eqn:entangled}) can be transferred to an entangled one
via Bell-state measurement on the orthogonal photon states
$|\pm\rangle$. For this we define the Bell bases as
$\psi^{\pm}=\frac{1}{2}\bigr(|+\rangle_{1L}|-\rangle_{2L}\pm|-\rangle_{1L}|+\rangle_{2L}\bigr)$
and
$\phi^{\pm}=\frac{1}{2}\bigr(|+\rangle_{1L}|+\rangle_{2L}\pm|-\rangle_{1L}|-\rangle_{2L}\bigr)$.
Using Bell-state measurements the eq. (\ref{eqn:entangled}) can be
recast into the following entangled states
\begin{eqnarray}\label{eqn:entangled2}
|\Psi\rangle^{\pm}_{12A}=\frac{1}{2}
\sqrt{N_+N'_-}|\frac{\bar\alpha}{\sqrt{2}}\rangle_1\otimes|\frac{-\bar\beta}{\sqrt{2}}\rangle_2
\pm\frac{1}{2}\sqrt{N'_+N_-}|\frac{-\bar\alpha}{\sqrt{2}}\rangle_1\otimes|\frac{\bar\beta}{\sqrt{2}}\rangle_2,
\end{eqnarray}
and
\begin{eqnarray}\label{eqn:entangled3}
|\Phi\rangle^{\pm}_{12A}=\frac{1}{2}
\sqrt{N_+N'_+}|\frac{\bar\alpha}{\sqrt{2}}\rangle_1\otimes|\frac{\bar\beta}{\sqrt{2}}\rangle_2
\pm\frac{1}{2}\sqrt{N'_-N_-}|\frac{-\bar\alpha}{\sqrt{2}}\rangle_1\otimes|\frac{-\bar\beta}{\sqrt{2}}\rangle_2,
\end{eqnarray}
respectively. The above results also show that the entanglement of
final states can be controlled with present EIT technique and
proper Bell-state measurements. Particularly, when $\alpha=\beta$,
the state $|\Psi\rangle^-_{12A}$ reduces into the maximum
entangled state. By now we successfully generate entangled states
of two-component spatially separated atom lasers, for which the
Schrodinger cat state is not needed for the input probe lights.
Noting that the Bell measurements on the photon states has been
widely studied \cite{Bell2}, the present scheme of spatially
separating the entangled atom lasers will be interesting for
quantum information processing.

Before conclusion, we would like to emphasize validity of the
approximations used in above derivation. Firstly, for the
adiabatic condition we have assumed the perfect two-photon
resonance and zero decay from the excited states. The
non-vanishing two-photon detunings can lead to an additional term
in the expansion of field $\Phi^l_{e_j}$:
\begin{eqnarray}\label{eqn:addition}
\hat\Phi^l_{e_j}\rightarrow\hat\Phi^l_{e_j}
+\frac{\Omega^2_{0j}(\delta_j+\mu_{jb}n)}{\Omega^2_{0j}-(\delta_j+\mu_{jb}n)(\Delta-i\gamma)}
\frac{g\hat{\cal E}_j}{\Omega_{0j}}\hat\Phi^l_b, \ \ j=1,2,
\end{eqnarray}
where the decay rate $\gamma$ from the excited states are
considered. Apparently the imaginary part in above formula can
lead to a loss in the solution of equation of probe fields: ${\cal
E}_j(z,t)\rightarrow e^{-\eta_j}{\cal E}_j(z,t)$ with $\eta_j>0$.
Here the calculation of factor $\eta_j$ is the same with that in
ref. \cite{6} and one can easily obtain
$\eta_j\leq|\delta_j+\mu_{jb}n|L/v_0$. Thus when
$\eta_j\leq|\delta_j+\mu_{jb}n|\frac{L}{v_0}\ll1$, the loss can be
safely neglected. As we know the residual Doppler shift of the
$|q_j\rangle\rightarrow |e_j\rangle$ transition can result in a
two-photon detuning through $\delta_j=\Delta
v(\vec{k}_{pj}-\vec{k}_{sj})\cdot\vec{e}_z(j=1,2)$, where
$\Delta{v}$ denotes the difference of the velocity in $z$
direction with respect to the resonant velocity class, present
condition also reads $\Delta
v/v_0\ll\min\{\frac{1}{|k_{p_j}-k_{s_j}|}(\frac{1}{L}-\frac{\mu_{bj}n}{v_0}),
j=1,2 \}$ and $\frac{v_0}{\mu_{bj}nL}>1$. The later condition is
easy to satisfied, because $\mu_{bj}n=4\pi n\hbar^2a_{bj}/m \
(a_{bj}\sim nm)$ is a much small factor. Another limitation of the
above discussion is set by the dephasing of the atom fields during
their propagation time $\Delta T$ from the entrance to the output
point. Obviously, $\Delta T$ can be chosen as small as (generally
larger than) the initial pulse length, and the compressed length
in the medium $\Delta L=V_{g_r}\Delta T$ should still sufficiently
larger than the wave length $\lambda$. Since the group velocity is
larger (or equal) than the propagation velocity $v_0$ of the
atomic beam ($V_{g_r}\rightarrow v_0$, when
$\Omega_0\rightarrow0$), the condition $\Delta L\gg\lambda$ can
easily be satisfied.

In conclusion we obtain two-component spatially separated
entangled atom lasers via quantum state transfer technique in a
five-level $M$-type system. The atom-atom collisions can yield an
effective Kerr susceptibility for this system and lead to the
self- and cross-phase modulation between the output atom/light
fields. This effect results in a large phase shift, which is
dependent on the factor $c/v_0$ and the space distribution of
Rabi-frequency $\Omega_{0j}(z)$ of the stokes fields, for the
output fields and has potential applications. Particularly,
considering the most ``classical", non-entangled coherent input
lights and under different conditions of space-dependent control
fields, we can obtain the entanglement of atom lasers and of
atom-light fields, etc. Furthermore, based on the Bell-state
measurement, an useful scheme is proposed to spatially separate
the generated entangled atom lasers in our paper. Under proper
Bell-state measurement, one can even obtain the maximum
entanglement of remote atom lasers. The large phase shift can also
be used to implement a quantum phase gate between two dark-state
polaritons (DSPs) \cite{5,6,liu} (whose input states are photons
and output states are atom lasers) when the input quantized probe
lights are both in the single-photon state.
\\

This work is supported by NSF of China under grants No.10275036
and No.10304020, and by NUS academic research Grant No. WBS:
R-144-000-071-305.



\noindent\\

\begin{figure}[ht]
\includegraphics[width=0.50\columnwidth]{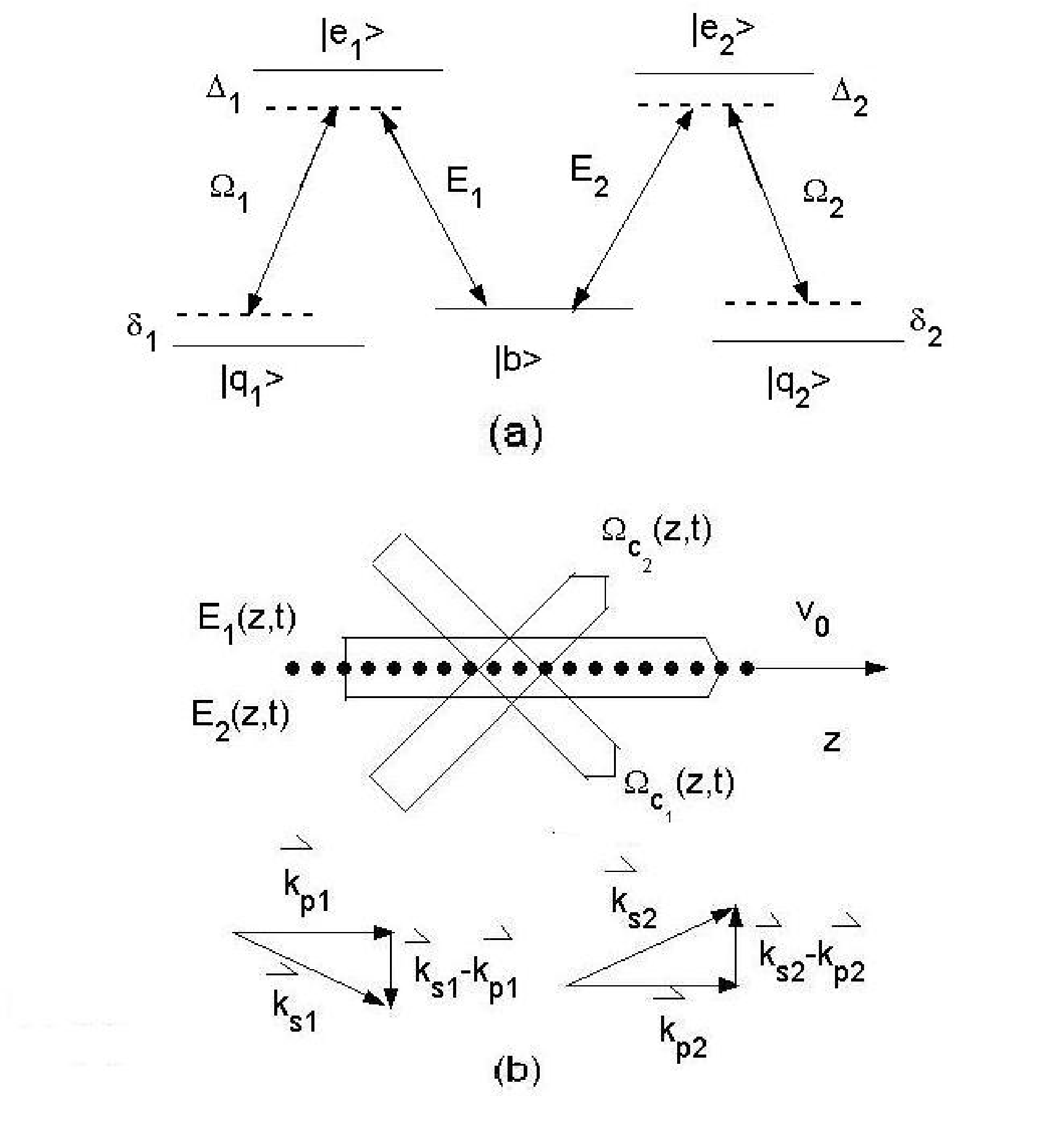}
\caption{(a)Beam of $M$ type atoms coupled to two control fields
and two quantized probe fields. (b)To minimize effect of
Doppler-broadening, geometry is chosen such that
$(\vec{k}_{pi}-\vec{k}_{si})\cdot\vec{e}_z\approx0$ $(i=1,2)$.}
\label{}
\end{figure}

\begin{figure}[ht]
\includegraphics[width=0.65\columnwidth]{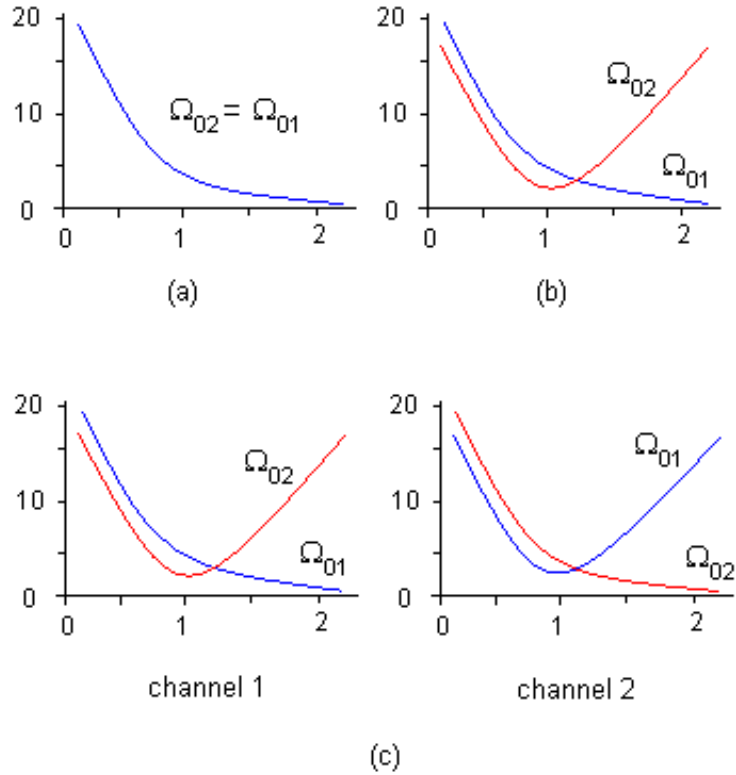}
\caption{(color online) Space-dependent Rabi-frequencies of
control fields. (a) the Rabi-frequencies $\Omega_{01}=\Omega_{02}$
so that $\theta_1(L)=\theta_2(L)=\pi/2$ at the output point. (b)
$\theta_1(L)=\pi/2, \theta_2(L)=0$ at the output point. (c) For
channel 1, $\theta_1(L)=\pi/2, \theta_2(L)=0$, whereas for channel
2, $\theta_1(L)=0, \theta_2(L)=\pi/2$.} \label{}
\end{figure}

\begin{figure}[ht]
\includegraphics[width=0.80\columnwidth]{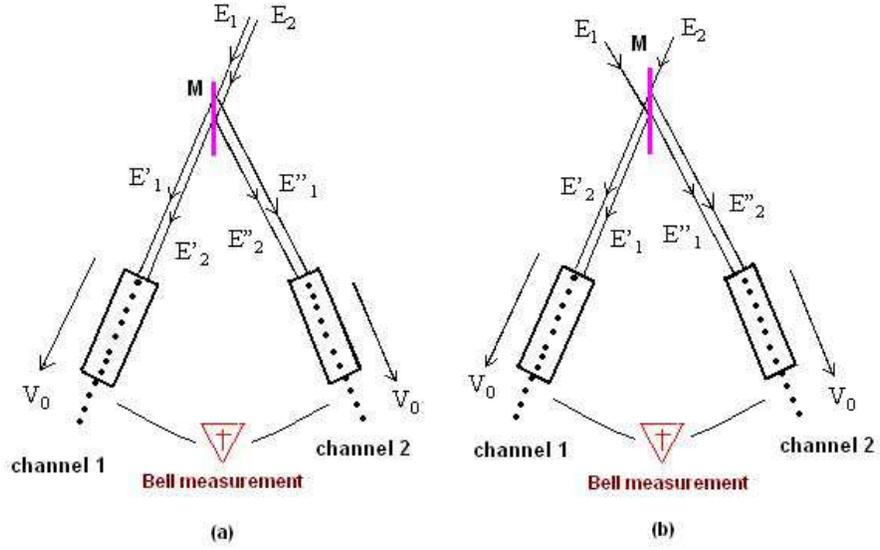}
\caption{(color online) (a)(b) The schematic set-up for generation
of two spatially separated entangled atom lasers under the
condition shown in Fig. 2(c). Based on the Bell measurement on the
photon states, the output atom lasers become entangled.} \label{}
\end{figure}

\end{document}